# Optimization of target film materials and protective coatings for sealed neutron generator

Yingying Cao [a], Sijia Zhou [a], Pingwei Sun [a], Jiayu Li [a], Shangrui Jiang [a], Shiwei Jing [a, *]

[a] *School of Physics, Northeast Normal University, Changchun, 130024, China*

*Corresponding author. E-mail address: jingsw504@nenu.edu.cn

**Abstract:** Magnesium target film has better thermal stability and neutron yield than titanium target, making it a potential neutron generator target film material. The radiation resistance of elemental magnesium targets is relatively weak, and their radiation resistance can be improved by alloying magnesium target films. The irradiation damage of pure magnesium targets and magnesium alloy target films was studied using SRIM. The results indicate that the irradiation damage of magnesium alloy target films (magnesium-niobium, magnesium-zirconium alloys) is lower than that of pure magnesium targets. In addition, under the same alloy ratio, the radiation resistance of magnesium-niobium alloy target film is better than that of magnesium-zirconium alloy. In order to further investigate the performance of magnesium alloy target films, the incident ion energy, protective coatings (nickel oxide, aluminum oxide, palladium oxide), magnesium alloy target films, and alloy doping ratios (0.2, 0.4, 0.6, 0.8, 1.0) were changed. After calculating the effects of the above conditions on the neutron generator yield, sputtering yield, and considering irradiation damage, it was determined that a magnesium-zirconium alloy with a doping rate of 0.2 and a nickel oxide protective coating with a thickness of 7.5 nm are potential target film materials for the neutron generator.

***Keywords***：Neutron generator; Alloy magnesium target; Irradiation; Neutron yield; SRIM

## 1. Introduction

Sealed neutron generator is widely used in multiple fields due to its compact size, portability, good monochromaticity, easy to control off, and capability to prevent radioactive pollution [1-5]. Neutron generator is a compact electro-vacuum device, composing ion source, target, acceleration system and pressure regulation system. The target assembly comprises the target film and the target substrate. The target film serves as the principal site for DD/DT fusion reactions, consequently exerting a substantial impact on the neutron yield, operational lifespan, and thermal stability of the neutron generator. Hydrogen absorption target film materials, including titanium, zirconium, scandium, erbium, yttrium, and lanthanum, are commonly used as target film materials. During the transportation and storage processes, a dense oxide layer develops on the metal surface, leading to a reduction in neutron yield as the thickness of the oxide layer increases [6]. Deuterium ion beams with varying energies continuously bombard the target membrane surface, leading to irradiation damage. This damage induces defects in the target, affecting the neutron yield. Furthermore, the ion source emits impurity ions such as $C^+$, $N^+$, and $O^+$, which are mixed in the deuterium ion beam. These impurities can cause damage or even ablation to the ion-sputtered target surface, thereby affecting the target life.

Significant progress has been achieved in enhancing the performance of neutron generators.

Huang et al. [7] selected pure molybdenum, characterized by their superior mechanical strength and better thermal conductivity, as the target film. The neutron yield reach $2\times10^8$ n/s. Guo et al. [8] utilized SRIM simulations to calculate the neutron yields for scandium-titanium, molybdenum-titanium and niobium-titanium alloy target films with ratios of 0.2, 0.4, 0.6, 0.8, and 1.0. The results indicated that the alloy ratio of 0.2 scandium-titanium neutron yield is highest. Alloying the target effectively prevented the formation of an oxide layer on the target surface, enhanced the mechanical properties of titanium, mitigated hydrogen embrittlement in the target film, and reduced damage from impurity ions to the target. However, alloying titanium targets also led to an increase in atomic number, enhance deuterium ions stopping power, and consequently lower neutron yields. Zhang et al. [9] through SRIM simulation, it has been demonstrated that the neutron yield of the magnesium target film exceeds that of the titanium target. This is attributed to the magnesium target has a higher hydrogen release temperature, greater hydrogen absorption and weak ion stopping power. However, the irradiation resistance of the magnesium target is weak, and the experimental results show that the neutron yield of the magnesium target film decay with time, while the neutron yield of the titanium target does not decay with time. In order to avoid energy loss due to the formation of an oxide layer on the surface of the target film and to prevent target damage from impurity ions (such as $C^+$, $N^+$, $O^+$, etc.) in the beam during ion sputtering, researchers applied a protection coat on the target film's surface. This approach aims to minimize surface ablation of the target film and reduce helium release resulting from tritium decay within the film. Falabella et al. [10] deposited a palladium layer with a thickness of 7.5-10 nm on the surface of a titanium film, demonstrating that this palladium layer can effectively prevent the target film from being oxidized. A. M. Zakharov et al. [11] by modifying the oxide layer of the surface layer of the titanium target film, reducing the tritium loss and improve the service life of the neutron generator.

    In this study, in order to enhance the irradiation resistance of magnesium target film, niobium and zirconium were doped into magnesium metal to form alloy magnesium target film materials. The irradiation damage to the target films of magnesium-niobium and magnesium-zirconium alloy was evaluated using SRIM simulations. To further investigate the properties of magnesium-niobium and magnesium-zirconium alloy, the neutron yield and sputtering yield of various alloy ratios were calculated using SRIM. In order that prevent energy loss from oxide formation on the surface of the magnesium alloy, and impurity ions present in the deuterium beam flow can sputter the target surface, thereby reducing the life of the neutron generator. Nickel oxide, alumina oxide and palladium oxide were selected as protective coatings. The neutron yield, sputtering yield of the magnesium-niobium and magnesium-zirconium alloy target films are simulated when these three oxides serve as protective coatings. The research results can be used as a reference for selecting the target film and the protective coatings material of the neutron generator.

## 2. Irradiation damage

    The irradiation resistance of the target film material is correlated with the displacement threshold energy, the higher the threshold value, the stronger the irradiation resistance. The incident particles on the target film collide with its atoms, transferring a fraction of their energy to these atoms. This energy transfer increases the energy states of the target atoms, leading to primary collision atoms. When the transfer energy exceeds the displacement threshold will secondary collision atoms

be ejected. The displaced atoms will then continue to collide, resulting in a cascade of collisions and the formation of defects. The incident particle type, energy, and irradiation dose all influence the extent of irradiation damage. Consequently, the concept of damage dose, measured in Displacements Per Atom (DPA), is introduced to quantify the average number of times each lattice atom is displaced from its original position. DPA value of 1 indicates that each lattice atom has been displaced once on average. Taking the D-T neutron generator as an example, Equation (1) is used to calculate the irradiation damage of target film [12], as follows:

Irradiation damage (DPA) = vacancy×irradiation dose / target film atom density     (1)

where the vacancy is determined by the energy of the incident particle the displacement threshold energy, etc. The vacancy can be calculated using the SRIM software. Displacement threshold energy is given in **Table 1**.

Table 1 The displacement threshold energy of element [13]

| Element | Mg | Nb | Zr |
|---|---|---|---|
| Displacement threshold energy / eV | 20 | 78 | 40 |

## 2.1 Irradiation damage with pure magnesium targe film

The irradiation damage on the pure magnesium target film was calculated when the deuterium ion beam current was 300 μA and the incident energy was 120 keV. The results are shown in **Fig. 1**. The results indicate that the deuterium ion incident depth of 1.5 μm achieves a peak displacement damage of 19.2 DPA. Integral to **Fig. 1**, total irradiation damage value for the pure magnesium target film is $9.21\times10^4$ DPA.

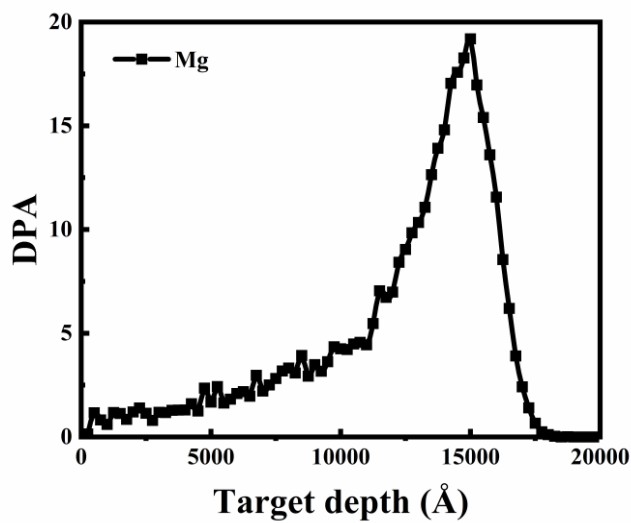

Fig. 1. The DPA depth distribution of the magnesium target film.

## 2.2 Irradiation damage with magnesium alloy target film

Radiation damage can lead to atomic displacement within the target film, resulting in tritium loss, decreased neutron yield, and consequently impacting the performance of the neutron generator.

In order to enhance irradiation resistance, the magnesium target film is doped with varying ratios of niobium, zirconium to form the magnesium alloy target film. There are two primary reasons for selecting these two metal elements. Firstly, niobium exhibits excellent sputtering resistance, while zirconium possesses a low thermal neutron cross section and superior corrosion resistance, making both materials important candidates for the target film in neutron generators, and scholars have carried out a lot of research on magnesium-niobium and magnesium-zirconium alloy [14-15]. Secondly, the displacement threshold energy of niobium and zirconium is higher compared to that of magnesium, which is beneficial to improve the irradiation resistance of the target film material.

The irradiation damage was calculated for the neutron generator with magnesium alloy target film when the deuterium ion beam current was 300 μA, the incident energy was 120 keV. The alloy target films include magnesium-niobium alloys and magnesium-zirconium alloys with atomic ratios of 0.2, 0.4, 0.6, 0.8, 1.0, the results are shown in **Fig. 2**. It can be seen form **Fig. 2** that the displacement peak of both the magnesium-niobium alloy and the magnesium-zirconium alloy decreases as the doping ratio increases. Additionally, under the same ratio, the displacement peak of the magnesium-zirconium alloy target film is higher than that of the magnesium-niobium alloy.

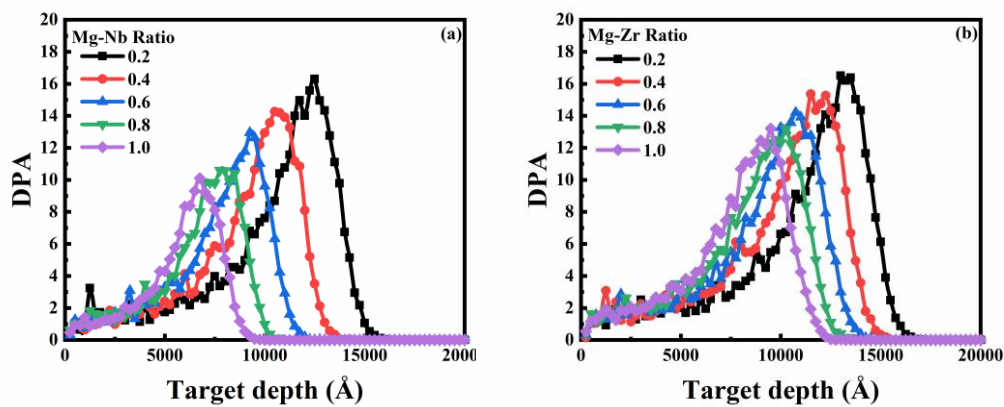

**Fig. 2.** The DPA depth distribution of target film: (a) Mg-Nb alloy (b) Mg-Zr alloy.

The **Fig. 2a** and **Fig. 2b** are integrated respectively, the results demonstrate that the addition of niobium and zirconium element both enhance the irradiation resistance of the magnesium target film. From a microstructural perspective, the atomic radii of niobium and zirconium differ significantly from that of magnesium. The lattice distortion induced during the alloying process can effectively impede dislocation motion and mitigate the migration and aggregation of irradiation-induced defects, thereby reducing the extent of irradiation damage. Moreover, alloying may promote the precipitation of secondary phases. These finely dispersed secondary-phase particles serve as pinning points at grain boundaries and within grains, further enhancing the material's irradiation resistance and stability. And the irradiation resistance of the magnesium-niobium alloy target film is superior to that of the magnesium-zirconium. The total irradiation resistance of the magnesium-niobium alloy target film with an alloy ratio of 0.2-1.0 has increased by 15.9 %, 31.8 %, 44.0 %, 59.4 %, 67.6 % compared to the pure magnesium, The total irradiation damage of magnesium-zirconium alloy with 0.2-1.0 alloy increased by 7.6 %, 13.6 %, 24.4 %, 32.2 %, 39.4 %, respectively.

# 3. Neutron yield

For the D-T neutron generator, the neutron yield is described by Equation 2 [16]:

$$Y = \frac{AR \cdot I \cdot N_T}{e} \sum_{k=1}^{3} k \cdot f_k \int_0^{E/k} \frac{\sigma(E)}{-S(E)} dE \qquad (2)$$

Where Y(E) represents the neutron yield from the D-T reaction, AR denotes the ratio of tritium ions in the target film, I is the incident deuterium ion beam current, and $N_T$ indicates the density of deuterium ions in the target film. The D-T reaction cross section, denoted as σ(E), is illustrated in **Fig. 3**. It is evident that the reaction cross section reaches its maximum value when the beam energy is 110 keV. The stopping power of deuterium ions, S(E), can be simulated using the SRIM software. E represents the energy of deuterium ions incident perpendicular to the target film surface. The indices k = 1, 2, and 3 correspond to monatomic, diatomic, and triatomic deuterium ions, respectively. And $f_k$ is a ratio of certain ion. However, the energy per atom of triatomic deuterium ions is lower than the minimum energy monitored by the neutron detection system [17], and their proportion is minimal. Therefore, the contribution of triatomic deuterium ions to the neutron yield is considered negligible.

An oxide layer forms on the surface of the target film during the transportation and storage of the neutron generator. Notably, according to reference [18], the thickness of this oxide layer on the magnesium target is 250 nm. This study assumes that the oxide layer is composed of a uniform magnesium oxide. Energy loss is experienced as the incident particle traverses the oxide layer in a direction perpendicular to the surface of the target film. Based on these assumptions, Equation (2) can be transformed into Equation (3):

$$Y = \frac{AR \cdot I \cdot N_T}{e} \sum_{k=1}^{3} k \cdot f_k \cdot T(E) \int_0^{E/k} \frac{\sigma(E)}{-S(E)} dE \qquad (3)$$

where T(E) is the transmittance of incident ions with varying energies through MgO.

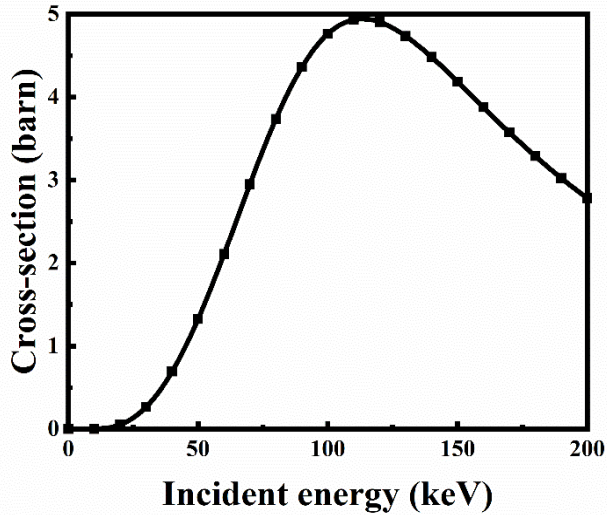

**Fig. 3.** D-T nuclear reaction cross section

## 3.1 Magnesium alloy target film

The SRIM software was utilized to simulate the energy loss and neutron yield of deuterium ions under a beam current of 300 μA and incident energy ranging from 40 to 150 keV. The simulation results are presented in **Fig. 4** and **Fig. 5.**

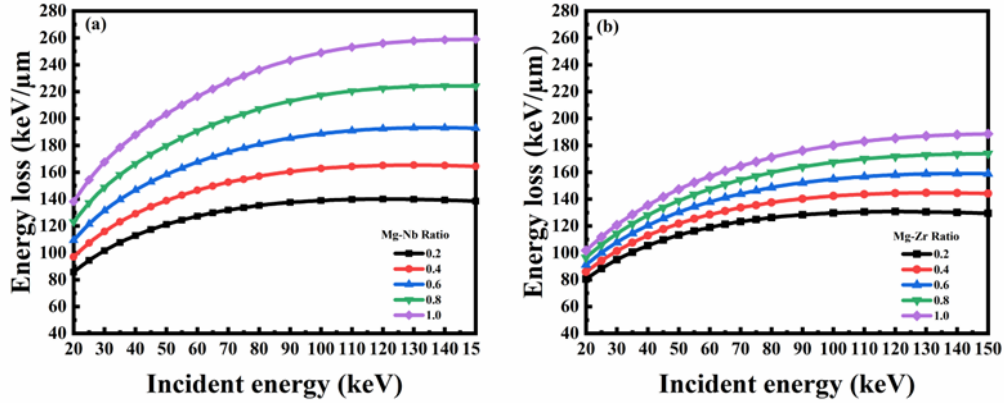

**Fig. 4.** The Stopping power of D$^+$ incident magnesium alloy target film (a) Mg-Nb alloy (b) Mg-Zr alloy.

As illustrated in **Fig. 4**, the ion stopping power of the magnesium-niobium alloy exceeds that of the magnesium-zirconium alloy under identical alloy compositions. Under the same alloy target film and incident energy, an increase in ratio of the doping metal results in enhanced stopping power of the magnesium alloy target film. This phenomenon can be attributed to the fact that elements with higher atomic numbers in the target film possess greater stopping power, which is consistent with the simulation results by Yang [19]. As can be seen from **Fig. 5**, the neutron yield increases with the increase of incident ion energy. In the same alloy target film and incident energy, the neutron yield decreases as the doping ratio increases. The maximum neutron yield is observed for a doping ratio of 0.2. Specifically, when the incident deuterium ion energy is 120 keV, the neutron yield for the magnesium-niobium alloy target film is $5.13 \times 10^9$ n/s, while that for the magnesium-zirconium alloy target film is $5.25 \times 10^9$ n/s. In the above case, 0.2 doping ratio magnesium-zirconium alloy target film has the highest neutron yield.

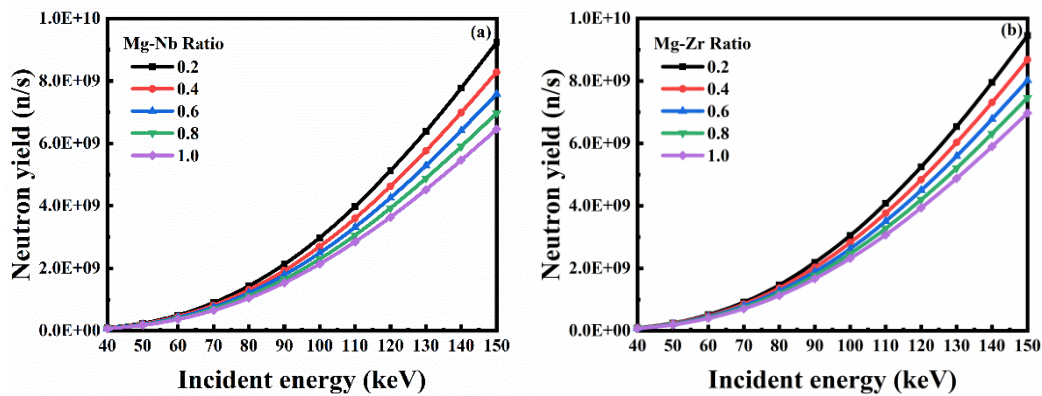

**Fig. 5.** The neutron yield with different incident ion energy of alloy target film (a) Mg-Nb alloy (b) Mg-Zr alloy.

## 3.2 Protective coatings on the magnesium alloy target film

The oxide layer that forms on the surface of the target film reduces the energy of deuterium ions as they enter the target, reducing the neutron yield. A thin protection overcoat can be applied to the surface of the target film. Materials such as nickel, aluminum, and palladium can serve as effective protection overcoats [20], preventing oxidation of the target film, minimizing the energy loss of deuterium ions as they pass through the surface layer, and consequently enhancing the neutron yield. Considering the practical problems, the oxide layers of nickel, aluminum, and palladium are selected as protective coatings for the magnesium alloy target film.

**Fig. 6** illustrates the transmittance of deuterium ions within an incident energy range of 40-150 keV as they traverse layers of 250 nm magnesium oxide, 7.5 nm nickel oxide, 7.5 nm alumina oxide, and 7.5 nm palladium oxide. The figure demonstrates that the transmittance increases as incident ion energy increase. Notably, the transmittance through nickel oxide, alumina oxide, and palladium oxide is significantly greater than that through magnesium oxide, and nickel oxide exhibiting the highest transmittance.

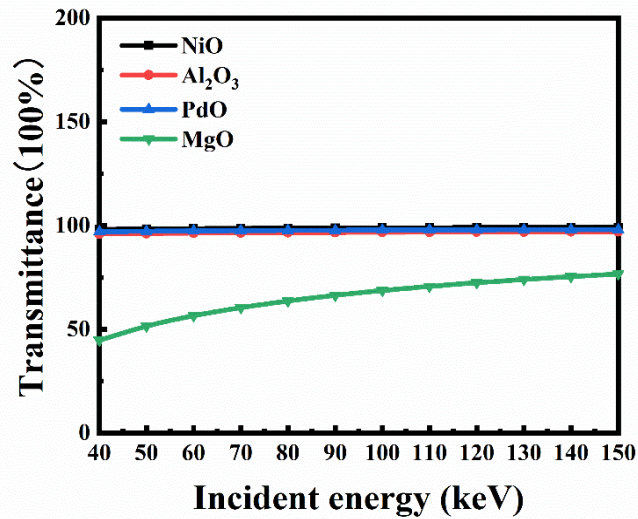

**Fig. 6.** The transmittance at 40-150 keV deuterium ion incident 7.5 nm nickel oxide, 7.5 nm aluminum oxide, 7.5 nm palladium oxide and 250 nm magnesium oxide

Meanwhile, the neutron yield was simulated at different ratios magnesium-niobium, magnesium-zirconium alloy target film with 250 nm magnesium oxide layer and with 7.5 nm nickel oxide, alumina oxide, palladium oxide protection overcoats which is bombarded by energy range of 40-150 keV deuterium, the results are shown in **Fig. 7** and **Fig. 8**. The results show that the neutron yield of the target film surface coated with the three protective coatings is significantly higher than that with the 250 nm thick magnesium oxide target film. Therefore, plating protective coatings on the surface of the target film can inhibit the formation of the protective oxide layer, and significantly enhancing the neutron yield of the neutron generator. Under identical incident conditions, the neutron yield of magnesium-niobium alloy and magnesium-zirconium alloy target films coated with nickel oxide is highest, followed by those coated with palladium oxide, while alumina oxide coated films exhibit the lowest yield. Under same alloy ratios, incident energy and protective coatings, the neutron yield of the magnesium-zirconium alloy target film remains superior to magnesium-niobium alloy. When incident deuterium ion energy is 120 keV, the 0.2 doping ratio magnesium-zirconium alloy target film with 7.5 nm nickel oxide protective coatings exhibit the highest neutron yield. The neutron yield is $5.09\times10^9$ n/s.

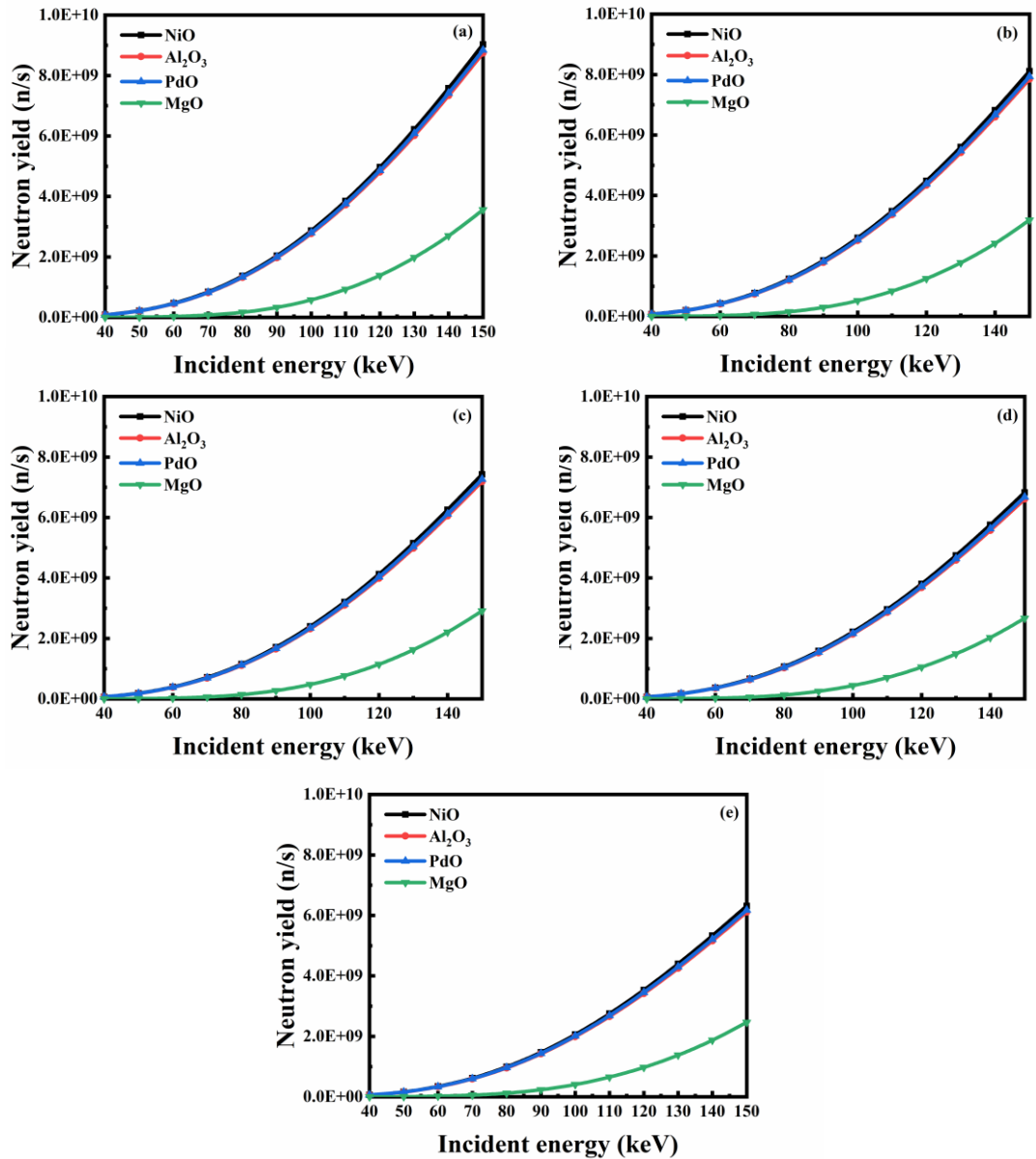

**Fig. 7.** The neutron yield of magnesium-niobium alloy target film with 250 nm MgO and 7.5 nm NiO, $Al_2O_3$ and PdO (a) 0.2 Mg-Nb alloy (b) 0.4 Mg-Nb alloy (c) 0.6 Mg-Nb alloy (d) 0.8 Mg-Nb alloy (e) 1.0 Mg-Nb alloy

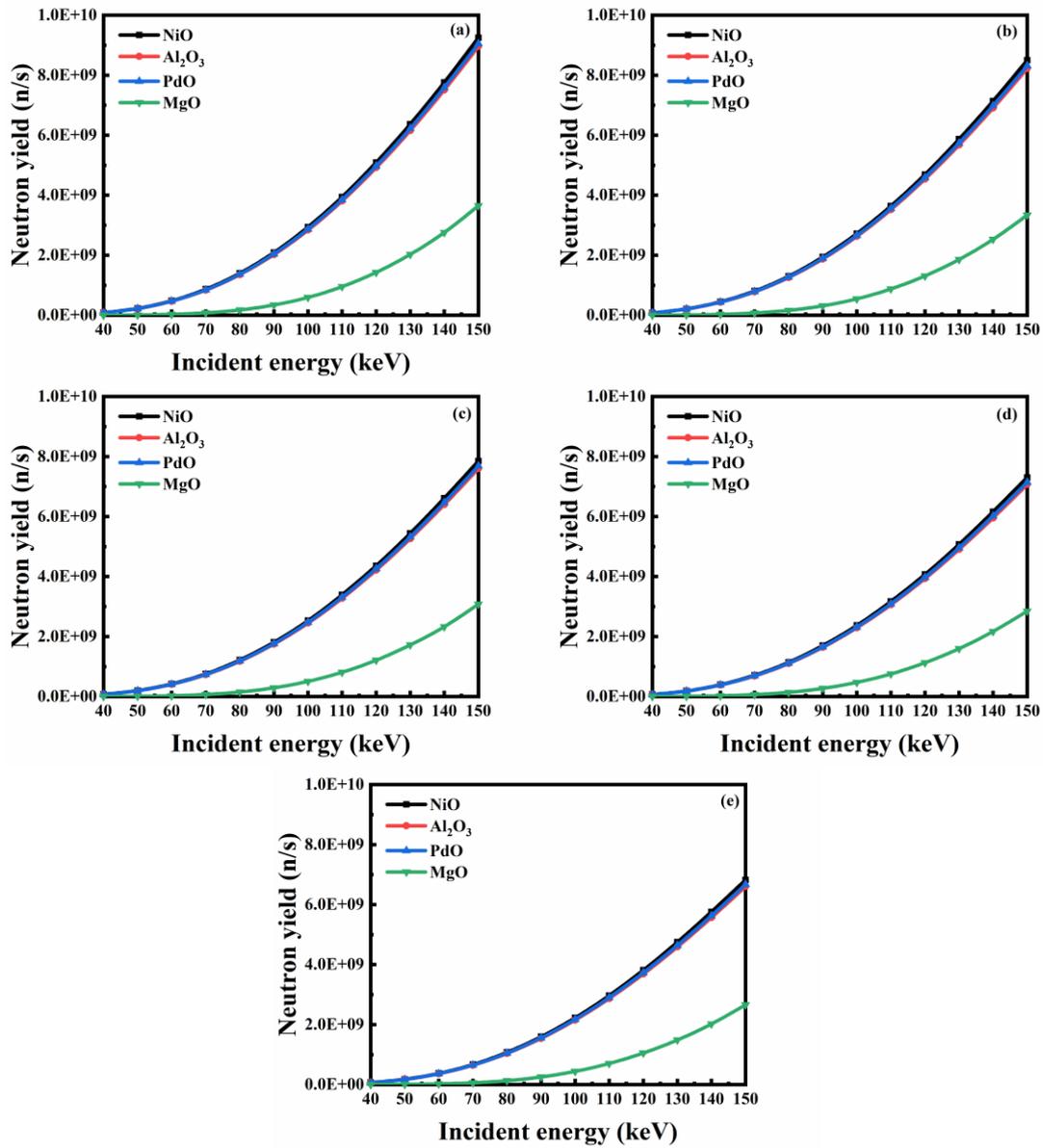

**Fig. 8.** The neutron yield of magnesium-zirconium alloy target film with 250 nm MgO and 7.5 nm NiO, Al$_2$O$_3$ and PdO (a) 0.2 Mg-Zr alloy (b) 0.4 Mg-Zr alloy (c) 0.6 Mg-Zr alloy (d) 0.8 Mg-Zr alloy (e) 1.0 Mg-Zr alloy.

## 4. Sputtering yield

There are many factors that influence the target life of a neutron generator, including irradiation damage on the surface caused by the deuterium ion beam, which leads to tritium release from the target film, as well as the sputtering effect of impurity ions such as C$^+$, N$^+$, and O$^+$ on the target film within the deuterium ion beam.

## 4.1 Sputtering yield of target film without protective coatings

The SRIM software was utilized to simulate the sputtering yields of $C^+$, $N^+$, and $O^+$ impurity ions with an incident energy of 120 keV in magnesium-niobium and magnesium-zirconium target films, with ratio ranging from 0.2 to 1.0. The simulation results are presented in **Fig. 9**. The results indicate that among the three impurity ions, $O^+$ exhibits the highest sputtering power, followed by $N^+$, with $C^+$ displaying the lowest sputtering power, are in agreement with the simulation by Yao et al. [21]. The alloy doping ratio significantly influences the sputtering yield of the target material. However, there is no linear relationship between the sputtering yield and the alloy doping ratio, indicating that an optimal alloy ratio exists to enhance the sputtering resistance of the alloy target film material. For the same alloy target film, the alloy with a doping ratio of 0.2 magnesium-niobium alloy target film demonstrates the strongest sputtering resistance, achieving a total sputtering yield of 1.72 atom·ion$^{-1}$. In contrast, the magnesium-niobium alloy with a doping ratio of 0.4 shows the weakest sputtering resistance, with a total sputtering yield of 2.30 atom·ion$^{-1}$. The doping ratio of 0.6 magnesium-zirconium alloy target film exhibits the lowest total sputtering yield of 1.80 atom·ion$^{-1}$, while a magnesium-zirconium alloy with a doping ratio of 0.2 exhibits the highest total sputtering yield of 2.55 atom·ion$^{-1}$.

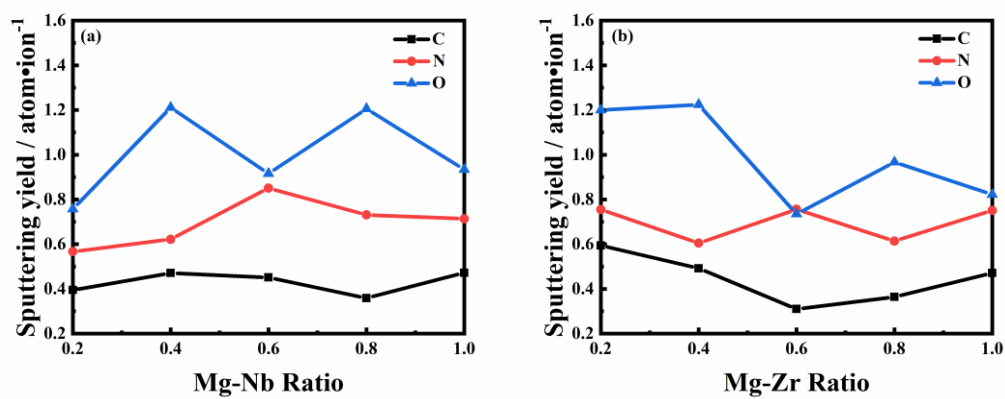

**Fig. 9** Relationship between sputtering yield and ratio of magnesium alloy (a) Mg-Nb alloy (b) Mg-Zr alloy.

## 4.2 Sputtering of target film with protective coatings

The impact of the protective layer on the anti-sputtering performance of the target film material was investigated. The SRIM software was utilized to simulate the sputtering yields of nickel oxide, alumina oxide, and palladium oxide under bombardment by 120 keV $C^+$, $N^+$, and $O^+$ impurity ions, within magnesium alloy ratio range of 0.2 to 1.0. As illustrated in **Fig. 10**, **Fig.11**, and **Fig.12**, the sputtering yields of magnesium alloy target film with alumina oxide protective coatings are the highest, indicating that the alumina oxide layer has the weakest capability to prevent impurity ions compared to the other two protective layers. The atomic sputtering yield of nickel oxide as magnesium alloy target film is the lowest, suggesting that the nickel oxide protective coating exhibits the strongest resistance among the three protective coatings. Therefore, nickel oxide demonstrates superior stability relative to the other materials used in the protective coatings. When nickel oxide is used as the protective coating, for the same alloy target film, the magnesium-niobium alloy target film with an alloy ratio of 0.2 has the lowest sputtering yield, which is 0.027 atom·ion$^{-1}$; the magnesium-zirconium alloy target film with an alloy ratio of 0.2 exhibits the strongest sputtering resistance, with a sputtering yield of 0.026 atom·ion$^{-1}$. It was determined that a magnesium-zirconium alloy with a doping rate of 0.2 and a nickel oxide protective coating with a thickness of 7.5 nm has strongest sputtering resistance.

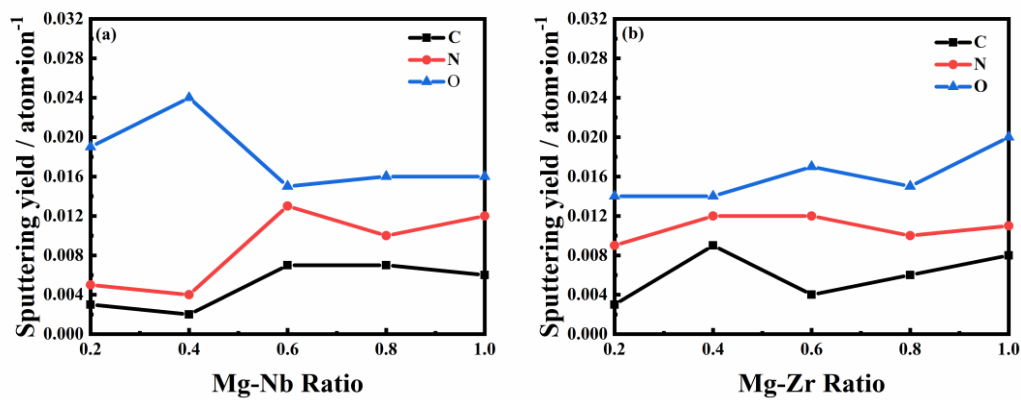

**Fig. 10** Relationship between sputtering yield and ratio of magnesium alloy with NiO protection overcoat (a) Mg-Nb alloy (b) Mg-Zr alloy.

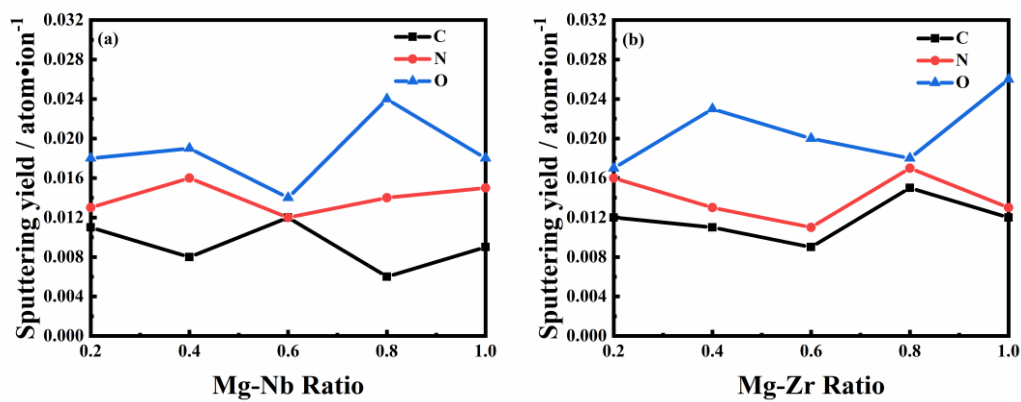

**Fig. 11** Relationship between sputtering yield and ratio of magnesium alloy with Al$_2$O$_3$ protection overcoat (a) Mg-Nb alloy (b) Mg-Zr alloy.

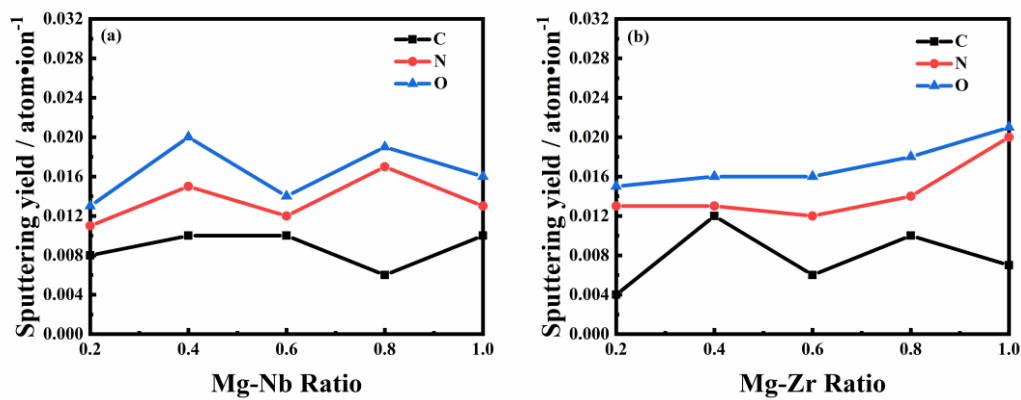

**Fig. 12** Relationship between sputtering yield and ratio of magnesium alloy with PdO protection overcoat (a) Mg-Nb alloy (b) Mg-Zr alloy.

# 5. Conclusion

This study aims to achieve long life and high yield neutron generator through the optimization of target film materials and surface layer materials. The irradiation damage of magnesium and two kinds of magnesium alloy target film was calculated, and the neutron yield and sputtering yield were simulated by changing magnesium alloy target materials and different protective coatings.

Through a systematic analysis of magnesium-based target films with different compositions under specific irradiation conditions, the result has elucidated the significant influence of alloying elements on the irradiation resistance performance of magnesium target films. Calculations indicate that niobium and zirconium, when used as alloying elements, can effectively mitigate the formation and accumulation of lattice defects during irradiation, thereby significantly enhancing the irradiation resistance of magnesium-based target films. Furthermore, under identical alloy compositions, the irradiation resistance of magnesium-niobium alloys surpasses that of magnesium-zirconium alloys. To further investigate the properties of the magnesium alloy target film, the SRIM software was utilized to simulate the neutron yield and sputtering yield for magnesium-niobium, magnesium-zirconium with different alloy ratios (0.2, 0.4, 0.6, 0.8, 1.0). The results indicate that at the same alloy ratio, the neutron yield of magnesium-zirconium alloy target film is better than magnesium-niobium alloy. When incident deuterium ion energy is 120 keV, magnesium-niobium alloy doping ratio of 0.2 exhibits the highest sputtering resistance, with a total sputtering yield of 1.72 atom·ion$^{-1}$. A protective coating is coated on the surface of magnesium alloy target film to enhance the neutron yield and improve sputtering resistance. By the calculation, when incident deuterium ion energy is 120 keV, the 0.2 doping ratio magnesium-zirconium alloy coated with a 7.5 nm NiO protective coating achieves a maximum neutron yield of $5.09 \times 10^9$ n/s and exhibits the highest sputtering resistance, with a total sputtering yield of 0.026 atom·ion$^{-1}$. Based on these conditions, the 0.2 doping ratio magnesium-zirconium alloy target film with NiO protective coating, is identified as a promising material combination for sealed neutron generators.

# CRediT authorship contribution statement

**Yingying Cao:** Investigation, data curation, methodology, software, validation, visualization, writing-original draft. **SiJia Zhou:** Software. **Pingwei Sun:** Validation. **Jiayu Li:** Visualization. **Shangrui Jiang:** Investigation. **Shiwei Jing:** Writing-review & editing, resources.

# Acknowledgments

This work was funded by the Department of Science and Technology of Jilin Province, China (20190303101SF).

# Reference

[1] Y.-M Song, H.-G Yang, J.-S Zhang, et al., The parameters test of a sealed D-T neutron tube, Journal of Isotopes 27(4) (2014) 199-202. http://doi org/10.7538/tws.2014.27.04.0199

[2] Farahmand M, Boston A J, Grint A N, et al., Detection of explosive substances by tomographic inspection using neutron and gamma-ray spectroscopy, Nuclear Instruments and Methods in Physics Research B, 261 (2007), 396-400. http://doi org/10.1016/j.nimb.2007.04.177

[3] M.-Y Yang, S. -L Fu, X, -M Chen, et al., Creation of wheat-rye small-fragment Translocation using Fast Neutron Irradiation, 34(5) (2014) 609-614. http://doi org /10.7606/j.issn.1009-1041.2014.05.06

[4] C.-B Lu, P. X, J.-X, et al., Preliminary experimental research of detail sensitivity in fast neutron photography, Nuclear Techniques, 38(8) (2015) 080202. http://doi org/10.11889/j.0253-3219.2015.hjs.38.08020

[5] Ahn S-K, Lee T-H, Shin H-S, et al., Simulation and preliminary experimental results for an active neutron counter using a neutron generator for a fissile material accounting, Nuclear Instruments and Methods in Physics Research A, 609 (2009) 205-212. http://doi org/ 10.1016/j.nima.2009.08.025

[6] W.-k Liu, X.-H Cao, S.-M, et al., Effects of the thickness of oxide layer on yield of neutron from deuterium-tritium reaction. Nuclear Techniques, 30(8) (2007) 665-667.

[7] Z.-W. Huang, X.-H. Bai, C.-Q. Liu, et al., Study on secondary electron suppression in compact D–D neutron generator, Nuclear Science and Techniques, 30 (2019). http://doi org/10.1007/s41365-019-0596-0

[8] W.-T. Guo, S.-J. Zhao, Z.-T. Yu, et al., Effect of target material on neutron output and sputtering yield of D-D neutron tube, Nuclear Instruments and Methods in Physics Research Section B: Beam Interactions with Materials and Atoms, 473 (2020) 48-54. http://doi org/10.1016/j.nimb.2020.04.017

[9] S.-W Zhang, S. C, D. Xiao, et al., Performance of Mg target for neutron generator under deuterium ion irradiation. Vacuum, 218 (2023) 112642. http://doi org/10.1016/j.vacuum.2023.112642

[10] Falabella S, Tang V, Ellsworth J L, et al., Protective overcoatings on thin-film titanium targets for neutron generator. Nuclear Instruments and Methods in Physics Research A, 736 (2014) 107−111. http://doi org/10.1016/j.nima.2013.10.045

[11] A. M. Zakharov, O. A. Dvoichenkova and A. E. Evisn., Modification of surface oxide layers of titanium target for increasing lifetime of neutron tube, Technology of Nuclear Materials, 78 (2015) 1643-1645. http://doi org/10.1134/S106377881514015X

[12] Ziegler J F, Ziegler M D, Biersack J P., SRIM – The stopping and range of ion matter, Nuclear Instruments and Methods in Physics Research B, 268 (2010) 1818-1823. http://doi org/10.1016/j.nimb.2010.02.091

[13] Kai Nordlund, Andrea E. Sand, Fredric Granberg, et al., Primary Radiation Damage in Materials: Review of Current Understanding and Proposed New Standard Displacement Damage Model to Incorporate in Cascade Defect Production Efficiency and Mixing Effects, NEA/NSC/DOC (2015)

[14] Konstantinas L, Eimutis J, Laurynas S, et al., Mg-Nb alloy films: Structure and stability in a balanced salt solution, Journal of Alloys and Compounds, 661 (2016) 322-330. http://doi org/10.1016/j.jallcom.2015.11.166

[15] R.-G Liu and W. Han., Molten − salt electrolytic preparation of magnesium − zirconium alloy, Journal of Yunnan University 23(6) (2014) 420-423. http://doi org/12.3969/j.issn1672-8513.2014.06.008

[16] Verbeke J M, Leung K N and Vujic J., Development of a sealed-accelerator-tube neutron generator, Appl. Radiation. Isotopes, 53 (2000) 801–809. http://doi org/10.1016/S0969-8043(00)00262-1

[17] X.-H Zhou, J. Lu, Y. Liu, et al., A concise method to calculate the target current ion species fraction in D-D and D-T neutron tube, Nuclear Instruments and Methods in Physics Research Section A, 987 (2021) 164836. http://doi org/10.1016/j.nima.2020.164836

[18] S.-W Zhang, Preparation and irradiation characteristic study of Mg-Based neutron source tritium target, Hefei, 2023.

[19] Z. Yang, J.-D Long, C.-H Lan, et al., Effects of ion source and target thickness on neutron yield from deuterium-tritium reaction neutron source, Nuclear Techniques 35(8) (2012) 591−595.

[20] W.-T. Guo, S.-J. Zhao, R.-X. Nian, et al., Impact of target material surface layer on neutron yield and target life of neutron tube, Radiation Physica and Chemistry, 186 (2021) 109548. http://doi org/10.1016/j.radphyschem.2021.109548

[21] Z.-E Yao, S.-W Chen, M.-Y Dong, et al., Sputtering Effect on Life-time of Titanium Tritide Target, Atomic Energy Science and Technology 37(1) (2003) 25−27.